\documentclass[usegraphicx]{mn2e}

\title[A Wave-Mechanical Approach to Cosmic Structure Formation]{A Wave-Mechanical Approach to Cosmic Structure
Formation}

\author[Coles \& Spencer]{Peter Coles and Kate Spencer\\
School of Physics \& Astronomy, University of Nottingham,
University Park, Nottingham, NG7 2RD, United Kingdom\\ }

\begin{document}

\maketitle

\begin{abstract}
The dynamical equations describing the evolution of a
self-gravitating fluid can be rewritten in the form of a
Schr\"{o}dinger equation coupled to a Poisson equation determining
the gravitational potential. This wave-mechanical representation
allows an approach to cosmological gravitational instability that
has numerous advantages over standard fluid-based methods. We
explore the usefulness of the Schr\"{o}dinger approach by applying
it to a number of simple examples of self-gravitating systems in
the weakly non-linear regime. We show that consistent description
of a cold self-gravitating fluid requires an extra ``quantum
pressure'' term to be added to the usual Schr\"{o}dinger equation
and we give examples of the effect of this term on the development
of gravitational instability. We also show how the simple wave
equation can be modified by the addition of a non-linear term to
incorporate the effects of gas pressure described by a polytropic
equation-of-state.
\end{abstract}


\section{Introduction}

The local Universe displays a rich hierarchical pattern of galaxy
clustering that encompasses a large range of length scales,
culminating in rich clusters and superclusters; the early
Universe, however, was almost smooth, with only slight ripples
seen in the cosmic microwave background radiation (e.g. Smoot et
al. 1992). Models of the evolution of structure link these
observations by appealing to the effect of gravitational
instability, e.g. Lahav et al. (2002). Low-amplitude primordial
perturbations of wave-like character within a largely homogeneous
universe become amplified by their own self-gravity, first
linearly which then, as their amplitudes build up, results in
non-linear collapse. The linear theory of perturbation growth for
cosmological density fluctuations is well-established, and is
founded on tried-and-tested representation of matter on
cosmological scales as a cold fluid (e.g. Peebles 1980). The
non-linear regime is much more complicated and generally not
amenable to analytic solution. Instead, numerical $N$--body
simulations using particles rather than waves to trace cosmic
structure, have led the way towards an understanding of strongly
developed clustering. Analytic approaches based on particles,
chiefly the Zel'dovich (1970) approximation, have also proved
useful. Although these agree to an extent with fully numerical
calculations (Coles, Melott \& Shandarin 1993), they break down
when particle trajectories cross to form singularities known as
caustics.

In this paper we adopt an approach to cosmic structure that has
advantages over both fluid and particle descriptions. Following
Widrow \& Kaiser (1993) and Coles (2002), we construct a formalism
in which the dynamics of gravitational is couched in the language
of quantum mechanics and governed by a Schr\"{o}dinger wave
equation. The advantages of this approach will be made apparent as
we explore its behaviour in simple examples which can be solved
exactly and, within the Sch\"{o}dinger approach, to various levels
of approximation. Our aim is to construct a useful tool for
evolving density fluctuations into the mildly non-linear regime,
beyond the breakdown of linear perturbation theory.

The importance of the search for analytic understanding of
structure formation is often overlooked in the light of the
tremendous advances that have been made recently in computational
astrophysics. One motivation for a wider range of theoretical
techniques is practical. High-resolution $N$-body experiments are
CPU-intensive and it is difficult to find sufficient resources to
use them to explore a large range of choices of initial
fluctuations, different cosmological parameters, and so on. A neat
analytic approximation can do this job more efficiently, at least
to the point of narrowing down the field of contenders for fuller
study. The other motivation is methodological: the complex
``cosmic web'' of large-scale structure requires {\em explanation}
rather than {\em reproduction}, and explanation rests on using
analytic approaches to identify the important physics as much as
possible.

The layout of this paper is as follows. In Section 2 we run
through the standard basics of gravitational instability in an
expanding universe, including perturbation theory and the
Zel'dovich approximation. We then, in Section 3, present the
details of the Schr\"{o}dinger approach and derive a
spherically-symmetric solution. In Section 4 we apply the method
to examples of one-dimensional collapse showing how various
approximations within the wave-mechanical framework behave when
compared to the real data. We sum up in Section 5.

\section{Cosmological Structure Formation}

\subsection{The Background Cosmology}
Space-times consistent with the Cosmological Principle can be
described by the Robertson--Walker metric
\begin{equation}
{\rm d}s^2 = c^2 {\rm d}t^2 - a^2(t)\left({{\rm d}r^2\over 1 -
\kappa r^2} + r^2 {\rm d}\theta^2 + r^2\sin^2\theta {\rm
d}\phi^2\right)  , \label{eq:l1a}
\end{equation}
where $\kappa$ is the spatial curvature, scaled so as to take the
values $0$ or $\pm 1$. The case $\kappa=0$ represents  flat space
sections, and the other two cases are  space sections of constant
positive or negative curvature, respectively. The time coordinate
$t$ is {\em cosmological proper time} and $a(t)$ is  the {\em
cosmic scale factor}. The dynamics of a Friedmann-Robertson-Walker
universe are determined by the Einstein gravitational field
equations which can be written
\begin{eqnarray}
3\left( \frac{\dot{a}}{a} \right)^{2} & = & 8\pi G\rho - {3\kappa
c^{2} \over a^2} + \Lambda c^2,\\ {\ddot{a}\over a} & = & - {4\pi
G\over 3} \left(\rho + 3 \frac{p}{c^2}\right) + {\Lambda c^2\over
3},
\\ \dot{\rho}& =& - 3 {\dot{a}\over a}\left(\rho + \frac{p}{c^2}
\right). \label{eq:l1b}
\end{eqnarray}
These equations  determine the time evolution of the cosmic scale
factor $a(t)$ (the dots denote derivatives with respect to
cosmological proper time $t$) and therefore describe the global
expansion or contraction of the universe. The behaviour of these
models can further be parametrized in terms of the Hubble
parameter $H=(\dot{a}/a)$ and the density parameter $\Omega=8\pi
G\rho/3H^2$, a suffix $0$ representing the value of these
quantities at the present epoch when $t=t_0$.

\subsection{Fluid Treatment}
 In the standard treatment of the Jeans Instability one
begins with the dynamical equations governing the behaviour of a
self-gravitating fluid. These are the {\em Euler equation}
\begin{equation}
{\partial ({\bf v})\over \partial t} + ({\bf v}\cdot{\bf
\nabla}){\bf v}  + {1\over \rho}{\bf \nabla} p + {\bf
\nabla}\phi=0~; \label{eq:Euler1}
\end{equation}
the {\em continuity equation} \begin{equation} {\partial\rho\over
\partial t} +  {\bf \nabla} (\rho{\bf v}) =
0~, \label{eq:continuity1}
\end{equation}
expressing the conservation of matter; and the {\em Poisson
equation}
\begin{equation} {\bf
\nabla}^2\phi = 4\pi G \rho~, \label{eq:Poisson1}
\end{equation}
describing Newtonian gravity. If the length scale of the
perturbations is smaller than the effective cosmological horizon
$d_H=c/H$, a Newtonian treatment of cosmic structure formation is
still expected to be valid in expanding world models. In an
expanding cosmological background, the Newtonian equations
governing the motion of gravitating particles can be written in
terms of
\begin{equation}
 {\bf x} \equiv {\bf r} / a(t)\end{equation} (the comoving spatial
coordinate, which is fixed for observers moving with the Hubble
expansion),
\begin{equation} {\bf v} \equiv \dot {{\bf r}} - H {\bf r} = a\dot
{{\bf x}}\end{equation} (the peculiar velocity field, representing
departures of the matter motion from pure Hubble expansion), $\rho
({\bf x}, t)$ (the matter density), and $\phi ({\bf x} , t)$ (the
peculiar Newtonian gravitational potential, i.e. the fluctuations
in potential with respect to the homogeneous background)
determined by the Poisson equation in the form
\begin{equation}
{\bf \nabla_x}^2\phi = 4\pi G a^2(\rho - \rho_0) = 4\pi
Ga^2\rho_0\delta. \label{eq:Poisson}
\end{equation}
In this equation and the following the suffix on $\nabla_x$
indicates derivatives with respect to the new comoving
coordinates. Here $\rho_0$ is the mean background density, and
\begin{equation}
\delta \equiv \frac{\rho-\rho_0}{\rho_0}
\end{equation}
is the {\em density contrast}. Using these variables the Euler
equation becomes
\begin{equation}
{\partial (a{\bf v})\over \partial t} + ({\bf v}\cdot{\bf
\nabla_x}){\bf v} = - {1\over \rho}{\bf \nabla_x} p - {\bf
\nabla_x}\phi~. \label{eq:Euler}
\end{equation}
The first term on the right-hand-side of equation (\ref{eq:Euler})
arises from pressure gradients, and is neglected in dust-dominated
cosmologies. Pressure effects may nevertheless be important in the
the (collisional) baryonic component of the mass distribution when
nonlinear structures eventually form. The second term on the
right-hand side of equation (\ref{eq:Euler}) is the peculiar
gravitational force, which can be written in terms of ${\bf g} =
-{\bf \nabla_x}\phi/a$, the peculiar gravitational acceleration of
the fluid element. If the velocity flow is irrotational, ${\bf v}$
can be rewritten in terms of a velocity potential $\phi_v$:
\begin{equation}{\bf v} = - {\bf
\nabla_x} \phi_v/a. \end{equation} This is expected to be the case
in the cosmological setting because (a) there are no sources of
vorticity in these equations and (b) vortical perturbation modes
decay with the expansion. We also have a revised form of the
continuity equation:
\begin{equation}
{\partial\rho\over \partial t} + 3H\rho + {1\over a} {\bf
\nabla_x} (\rho{\bf v}) = 0~. \label{eq:continuity}
\end{equation}
\subsection{Linear Perturbation Theory}
The procedure for handling perturbations is  to linearise the
Euler, continuity and Poisson equations by perturbing physical
quantities defined as functions of Eulerian coordinates, i.e.
relative to an unperturbed coordinate system. Expanding $\rho$,
${\bf v}$ and $\phi$ perturbatively and keeping only the
first-order terms in equations (\ref{eq:Euler}) and
(\ref{eq:continuity}) gives the linearised continuity equation:
\begin{equation}
{\partial\delta\over \partial t} = - {1\over a}{\bf \nabla_x}\cdot
{\bf v},
\end{equation}
which can be inverted, with a suitable choice of boundary
conditions, to yield
\begin{equation}
\delta = - {1\over a H f}\left({\bf \nabla_x}\cdot{\bf v}\right).
\label{eq:l9}
\end{equation}
The function $f\simeq \Omega_0^{0.6}$; this is simply a fitting
formula to the full solution (Peebles 1980). The linearised Euler
and Poisson equations are
\begin{equation}
{\partial {\bf v}\over\partial t} + {\dot a\over a}{\bf v} = -
{1\over \rho a}{\bf \nabla_x} p -{1\over a}{\bf \nabla_x}\phi,
\label{eq:l10}
\end{equation}
\begin{equation}
{\bf \nabla_x}^2\phi = 4\pi G a^2\rho_0\delta; \label{eq:l11}
\end{equation}
$|{\bf v}|, |\phi|, |\delta| \ll 1$ in equations (\ref{eq:l9}),
(\ref{eq:l10}) \& (\ref{eq:l11}). From these equations, and if one
ignores pressure forces, it is easy to obtain an equation for the
evolution of $\delta$:
\begin{equation}
\ddot\delta + 2H\dot\delta - {3\over 2}\Omega H^2\delta = 0.
\label{eq:perdel}
 \label{eq:l13b}
\end{equation}
For a spatially flat universe dominated by pressureless matter,
$\rho_0(t) = 1/6\pi Gt^2$ and equation (\ref{eq:l13b}) admits two
linearly independent power law solutions $\delta({\bf x},t) =
D_{\pm}(t)\delta({\bf x})$, where $D_+(t) \propto a(t) \propto
t^{2/3}$  is the growing  mode and $D_-(t) \propto t^{-1}$ is the
decaying mode.

The above considerations apply to the evolution of a single
Fourier mode of the density field $\delta({\bf x}, t) =
D_+(t)\delta({\bf x})$. What is more likely to be relevant,
however, is the case of a superposition of waves, resulting from
some kind of stochastic process in which the density field
consists of a  superposition of such modes with different
amplitudes. A statistical description of the initial perturbations
is therefore required, and any comparison between theory and
observations will also have to be statistical. Many versions of
the inflationary scenario for the very early universe (Guth 1981;
Guth \& Pi 1982; Linde 1982; Albrecht \& Steinhardt 1982) predict
the initial density fluctuations to take the form of a {\em
Gaussian random field} in which the initial Fourier modes of the
perturbation field have random phases.

The linearised equations of motion  provide an excellent
description of gravitational instability at very early times when
density fluctuations are still small ($\delta \ll 1$). The linear
regime of gravitational instability breaks down when $\delta$
becomes comparable to unity, marking the commencement of the {\it
quasi-linear} (or weakly non-linear) regime. During this regime
the density contrast may remain small ($\delta < 1$), but the
phases of the Fourier components $\delta_{\bf k}$ become
substantially different from their initial values resulting in the
gradual development of a non-Gaussian distribution function even
if the primordial density field is Gaussian owing to mode--mode
interactions.

Perturbation theory fails when non-linearity develops, but the
fluid treatment is intrinsically approximate anyway. A fuller
treatment of the problem requires a solution of the Boltzmann
equation for the full phase-space distribution of the system
$f({\bf x}, {\bf v}, t)$ coupled to the Poisson equation
(\ref{eq:Poisson1}) that determines the gravitational potential.
In cases where the matter component is collisionless, the
Boltzmann equation takes the form of a Vlasov equation:
\begin{equation}
{\partial f \over \partial t}= \sum_{i=1}^{3} \left({\partial
\phi\over
\partial x_i}{\partial f \over \partial v_i} - v_i {\partial f \over \partial
x_i}\right).
\end{equation}
The fluid approach  outline above can only describe cold material
where the velocity dispersion of particles is negligible. But even
if the dark matter is cold, there may be hot components of
baryonic material whose behaviour needs also to be understood.
Moreover, the fluid approach assumes the existence of a single
fluid velocity at every spatial position. It therefore fails when
orbits cross and multi-streaming generates a range of particle
velocities through a given point.

\subsection{The Zel'dovich Approximation}
Fortunately the formation of the main elements of large-scale
structure  can nevertheless be understood using relatively simple
tools, such as the Zel'dovich approximation (Zel'dovich 1970). Let
the initial (i.e. Lagrangian) coordinate of a particle in this
unperturbed distribution be ${\bf q}$. Now each particle is
subjected to a displacement corresponding to a density
perturbation. In the Zel'dovich approximation the  Eulerian
coordinate of the particle at time $t$ is  \begin{equation}{\bf r}
(t, {\bf q}) = a(t) [{\bf q}-b(t) {\bf \nabla}_{\bf q} \Phi_0
({\bf q})], \end{equation} where ${\bf r}=a(t){\bf x}$, with ${\bf
x}$ a comoving coordinate, and we have made $a(t)$ dimensionless
by dividing throughout by $a(t_i)$, where $t_i$ is some reference
time which we take to be the initial time. Notice that particles
in the Zel'dovich approximation execute a kind of inertial motion
on straight line trajectories. The derivative on the right hand
side is taken with respect to the Lagrangian coordinates. The
dimensionless function $b(t)$ describes the evolution of a
perturbation in the linear regime, with the
 condition $b(t_i)=0$, and therefore solves the equation
\begin{equation} \ddot {b} +2 \left({\dot a \over a}\right) \dot b- 4 \pi G \rho b =
0~;
\end{equation}
cf. equation (19). For a flat matter--dominated universe we have
 $b \propto t^{2/3}$ as before. The quantity
$\Phi_0({\bf  q})$ is proportional  to a velocity potential, of
the type introduced above, i.e. a quantity of which the velocity
field is the gradient:
\begin{equation}
{\bf V}= {d{\bf r} \over dt} - H {\bf r} = a {d {\bf x} \over dt}
= - a \dot b {\bf \nabla}_q \Phi_0({\bf q});
\end{equation} this means that the velocity field is irrotational. The
quantity $\Phi_0 ({\bf q})$ is related to the density perturbation
in the linear regime by the relation \begin{equation} \delta = b
\nabla^2 _{\bf q} \Phi_0, \end{equation} which is a simple
consequence of Poisson's equation.

The Zel'dovich approximation, though simple, has a number of
interesting properties. First, it is exact for the case of one
dimensional perturbations up to the moment of shell crossing. As
we have mentioned above, it also incorporates irrotational motion,
which is required to be the case if it is generated only by the
action  of gravity (due to the Kelvin circulation theorem). For
small displacements between ${\bf r}$ and $a(t){\bf q}$, one
recovers the usual (Eulerian) linear regime: in fact, equation
(22) defines a unique mapping between the coordinates ${\bf q}$
and ${\bf r}$ (as long as trajectories do not cross); this means
that
 $\rho ({\bf r} ,t) d^3 r =
\langle \rho(t_i) \rangle d^3 q$ or \begin{equation} \rho({\bf r},
t ) = { \langle \rho (t)\rangle \over \vert J ({\bf r}, t) \vert
}~,\end{equation} where $\vert J  ({\bf r}, t) \vert$ is the
determinant of the Jacobian of the mapping between ${\bf q}$ and
${\bf r}$: $\partial {\bf r} / \partial {\bf q}$.  Since the flow
is irrotational the matrix $J$ is symmetric and can therefore be
locally diagonalised. Hence \begin{equation}
 \rho({\bf r}, t ) =
\langle \rho (t)\rangle \prod_{i=1} ^3  [1 + b(t) \alpha_i({\bf
q})]^{-1}: \end{equation} the quantities $1 + b (t) \alpha_i$ are
the eigenvalues of the matrix $J$ (the $\alpha_i$ are the
eigenvalues of the deformation tensor). For times close to $t_i$,
when $\vert b(t) ~\alpha_i\vert \ll 1$, equation (26) yields
\begin{equation}
 \delta \simeq - (\alpha_1 + \alpha_2 + \alpha_3) b(t),
\end{equation}
which is just the law of perturbation growth obtained by solving
equation (\ref{eq:perdel}).

At some time $t_{sc}$, when $ b(t_{sc}) = - 1/\alpha_j$, a
singularity appears and the density becomes formally infinite in a
region where at least one of the eigenvalues (in this case
$\alpha_j$) is negative. This condition corresponds to the
situation where two points with different Lagrangian coordinates
end up at the same Eulerian coordinate. In other words, particle
trajectories have crossed and the mapping between Lagrangian and
Eulerian space is no longer unique. A region where the
shell--crossing occurs is called a caustic. For a fluid element to
be collapsing, at least one of the $\alpha_j$ must be negative. If
more than one is negative, then collapse will occur first along
the axis corresponding to the most negative eigenvalue. If there
is no special symmetry, one therefore expects collapse to be
generically one--dimensional, from three dimensions to two. We
shall make use of the fact that in one-dimensional cases, the
Zel'dovich solution is exact until the moment of shell-crossing.

The Zel'dovich approximation matches very well the evolution of
density perturbations in full $N$--body calculations until the
point where shell crossing occurs (Coles, Melott \& Shandarin
1993). After this, the approximation breaks down completely.
Particles continue to move through the caustic in the same
direction as they did before. Particles entering a pancake from
either side merely sail through it and pass out the opposite side.
The pancake therefore appears only instantaneously and is rapidly
smeared out. In reality, the matter in the caustic would feel the
strong gravity there and be pulled back towards it before it could
escape through the other side. Since the Zel'dovich approximation
is only kinematic it does not account for these close--range
forces and the behaviour in the strongly non--linear regime is
therefore described very poorly. Furthermore, this approximation
cannot describe the formation of shocks and phenomena associated
with pressure. Attempts to understand properties of non-linear
structure using a fluid description therefore resort to further
approximations (Sahni \& Coles 1995) to extend the approach beyond
shell-crossing. One method involves the so--called {\it adhesion
model} (Gurbatov, Saichev \& Shandarin 1989). In this  model one
assumes that the particles stick to each other when they enter a
caustic region because of an artificial viscosity which is
intended to simulate the action of strong gravitational effects
inside the overdensity forming there. This `sticking' results in a
cancellation of the component of the velocity of the particle
perpendicular to the caustic. If the caustic is two--dimensional,
the particles will move in its plane until they reach a
one--dimensional interface between two such planes. This would
then form a filament. Motion perpendicular to the filament would
be cancelled, and the particles will flow along it until a point
where two or more filaments intersect, thus forming a node. The
smaller is the viscosity term, the thinner will be the sheets and
filaments, and the more point--like will be the nodes. Outside
these structures, the Zel'dovich approximation is still valid to
high accuracy. See Shandarin \& Zel'dovich (1989) for further
discussion.

\subsection{The Strongly Nonlinear Regime}

Further into the non-linear regime, bound structures form. The
baryonic content of these objects may then become important
dynamically: hydrodynamical effects (e.g. shocks), star formation
and heating and cooling of gas all come into play. The spatial
distribution of galaxies may therefore be very different from the
distribution of the (dark) matter, even on large scales. Attempts
are only just being made to model some of these processes with
cosmological hydrodynamics codes, such as those based on Smoothed
Particle Hydrodynamics (SPH; Monaghan 1992), but it is some
measure of the difficulty of understanding the formation of
galaxies and clusters that most studies have only just begun to
attempt to include modelling the detailed physics of galaxy
formation. In the front rank of theoretical efforts in this area
are the so-called semi-analytical models which encode simple rules
for the formation of stars within a framework of merger trees that
allows the hierarchical nature of gravitational instability to be
explicitly taken into account.

The spatial distribution of particles obtained using the adhesion
approximation represents a sort of ``skeleton'' of the real
structure: non--linear evolution generically leads to the
formation of a quasi--cellular structure, which is a kind of
``tessellation'' of irregular polyhedra having pancakes for faces,
filaments for edges and nodes at the vertices . This skeleton,
however, evolves continuously as structures merge and disrupt each
other through tidal forces; gradually, as evolution proceeds, the
characteristic scale of the structures increases. In order to
interpret the observations we have already described, one can
think of the giant ``voids'' as being the regions internal to the
cells, while the cell nodes correspond to giant clusters of
galaxies. While analytical methods, such as the adhesion model,
are useful for mapping out the skeleton of structure formed during
the non--linear phase, they are not adequate for describing the
highly non--linear evolution within the densest clusters and
superclusters. In particular, the adhesion model cannot be used to
treat the process of merging and fragmentation of pancakes and
filaments due to their own (local) gravitational instabilities,
which must be done using full numerical computations.

\section{Wave Mechanics and Structure Formation}

\subsection{Introduction}
The combination of Eulerian fluid-based perturbation theory,
Lagrangian approaches like the Zel'dovich approximation and
brute-force $N$-body numerics has been successful at establishing
a standard framework within which the basic features of
large-scale structure can be described and understood. There
remain significant obstacles to fuller analytic description. The
three most interesting from the perspective followed in this paper
are:

\begin{enumerate}
\item {\bf Perturbation Theory.} Standard perturbation methods do
not guarantee a density field that is everywhere positive. This
basic problem is obvious in the standard case where there is a
Gaussian distribution of initial fluctuations. When the variance
of the distribution of $\delta$ is of order unity (its mean is, by
definition, always unity) then a Gaussian distribution assigns a
non-zero probability to regions with $\delta<-1$, i.e. with
$\rho<0$.
\item {\bf Shell-crossing.} Methods such as the Zel'dovich
approximation and its variations break down at shell-crossing,
where the density becomes infinite.
\item {\bf Gas Pressure.} Analytic techniques for modelling the
effects of gas pressure are scarce, even in relatively simple
systems such as the Lyman-$\alpha$ absorbing systems (Matarrese \&
Mohayee 2002). Fully numerical techniques, such as SPH, are
generally expected to be the way forward but even they are not yet
able to incorporate all relevant gas physics.\end{enumerate}

\subsection{The Widrow--Kaiser Approach}

A novel approach to the study of collisionless matter, with
applications to structure formation, was suggested by Widrow \&
Kaiser (1993). It involves re-writing of the fluid equations given
in Section 2 in the form of a non-linear Schr\"{o}dinger equation.
The equivalence between this and the fluid approach has been known
for some time; see Spiegel (1980) for historical comments.
Originally the interest was to find a fluid interpretation of
quantum mechanical effects, but in this context we shall use it to
describe an entirely classical system.

To begin with, consider the continuity equation and Euler equation
for a curl-free flow in which ${\bf v}=\nabla \phi)$, in response
to some general potential $V$. In this case the continuity
equation can be written
\begin{equation}
\frac{\partial \rho}{\partial t} + \nabla \cdot (\rho\nabla\phi)=
0~. \end{equation} It is convenient to take the first integral of
the Euler equation, giving
\begin{equation}
\frac{\partial \phi}{\partial t} + \frac{1}{2} (\nabla \phi)^2 =
-V~,
\end{equation}
which is usually known as the Bernoulli equation. The trick then
is to make a transformation of the form
\begin{equation}
\psi=\alpha \exp(i\phi/\nu)~,\label{eq:conpsi}
\end{equation}
where $\rho=\psi\psi^*=|\psi|^2=\alpha^2$; the wavefunction
$\psi({\bf x}, t)$ evidently complex. Notice that the dimensions
of $\nu$ are the same as $\phi$, namely $[L^2T^{-1}]$. After some
algebra it emerges that the two equations above can be written in
one equation of the form
\begin{equation}
i\nu \frac{\partial \psi}{\partial t} = -\frac{\nu^2}{2} \nabla^2
\psi + V\psi + P\psi~, \label{eq:scrap}
\end{equation}
where \begin{equation} P=\frac{\nu^2}{2} \frac{\nabla^2
\alpha}{\alpha}. \end{equation} To accommodate gravity we need to
couple  equation (\ref{eq:scrap}) to the Poisson equation in the
form
\begin{equation}
\nabla^2\Phi=4\pi G\psi \psi^*~,
\end{equation}
taking $V$ to be $\Phi$.

This system looks very similar to a Schr\"{o}dinger equation,
except for the extra ``operator'' $P$. The role of this term
becomes clearer if one leaves it out of equation (\ref{eq:scrap})
and works backwards. The result is an extra term in the Bernouilli
equation that resembles a pressure gradient. This is often called
the ``quantum pressure'' that arises when one tries to understand
a quantum system in terms of a classical fluid behaviour. Leaving
it out to model a collisionless fluid can be justified only if
$\alpha$ varies only slowly on the scales of interest. On the
other hand one can model situations in which one wishes to model
genuine effects of pressure by adjusting (or omitting) this term
in the wave equation. Widrow \& Kaiser (1993) advocated simply
writing
\begin{equation}
i\hbar {\partial \psi\over \partial t}= - {\hbar^2 \over 2m}
\nabla^2\psi + m \phi({\bf x})\psi,\label{eq:schrod}
\end{equation}
i.e. simply ignoring the quantum pressure term. In this spirit,
the constant $\hbar$ is taken to be an adjustable parameter that
controls the spatial resolution $\lambda$ through a de Broglie
relation $\lambda=\hbar/mv$. In terms of the parameter $\nu$ used
above, we have $\nu=\hbar/m$, giving the correct dimensions for
Planck's constant. Note that the wavefunction $\psi$ encodes the
velocity part of phase space in its argument through the ansatz
\begin{equation}
\psi({\bf x})=\sqrt{\rho({\bf x})} \exp [i\theta({\bf x})/\hbar],
\label{eq:simans}
\end{equation}
where $\nabla \theta({\bf x})={\bf p}({\bf x})$, the local
`momentum field'. This formalism thus yields an elegant
description of both the density and velocity fields in a single
function.

It should be clear how this approach bypasses immediately the
first two items of the list given in Section 3.1. First, the
construction (\ref{eq:conpsi}) requires that $\rho=|\psi|^2$ is
positive. Secondly, since particles are not treated as point-like
entities with definite trajectories, shell-crossing does not have
the catastrophic from in this approach that it does in the
Zel'dovich approximation.  Note that no singularities occur in the
wavefunction at any time. Although the simplest {\em ansatz}
(\ref{eq:simans}) does assume a unique velocity at every fluid
location, it is possible to construct more complex representations
that allow for multi-streaming (Widrow \& Kaiser 1993) although we
shall not use them in this paper.

\subsection{A Spherically-Symmetric Solution}

In order to illustrate how the wave equation approach maps onto a
fluid description, it is instructive to examine a familiar
problem: a static spherically symmetric self-gravitating cloud in
hydrostatic balance. This also allows us to explain how item (3)
of the list in Section 3.1 can be addressed. We  model the
pressure gradients needed to hold an object in equilibrium against
its self-gravity by  the addition to the potential of a term of
the form $\kappa^2 |\psi|^2$ (with $\kappa$ an
appropriately-chosen constant), which converts the original
equation (\ref{eq:schrod}) into a nonlinear Schr\"{o}dinger
equation:
\begin{equation}
i\nu {\partial \psi\over \partial t}= - {\nu^2 \over 2}
\nabla^2\psi + \Phi ({\bf x})\psi + \kappa^2 |\psi|^2 \psi + P\psi
\label{eq:schrodnl}
\end{equation}
(Sulem \& Sulem 1999), where the final term is the quantum
pressure. We will now look for solutions of the form
\begin{equation}
\psi=R\exp\left(\frac{i\phi}{\nu}\right),
\end{equation}
but we are looking for a static solution so obviously the fluid
flow velocity should be zero. We therefore need to set $\phi$ to
be constant in the above transformation so that $\nabla\phi$ can
be zero everywhere. Without loss of generality we can exploit the
global U(1) symmetry of quantum mechanics to set the phase to
zero. Hence we can set $\psi=R$ and consequently
$\nabla^2\psi=\nabla^2R$. The quantum pressure term has the form
\begin{equation}
P = \frac{\nu^2}{2} \frac{\nabla^2 R}{R}~,
\end{equation}
so that this now cancels the first term on the right-hand-side of
equation (\ref{eq:schrodnl}). In this event, and requiring a
static solution for $\psi$ we get
\begin{equation}
\Phi\psi + \kappa^2 |\psi|^2 \psi=0
\end{equation}
to be solved simultaneous with Poisson's equation. We therefore
obtain a simple Helmholtz equation for $\Phi$:
\begin{equation}
\nabla^2 \Phi +\left(\frac{4\pi G}{\kappa^2}\right) \Phi=0.
\label{eq:neat}
\end{equation}
The solution for spherically symmetry is straightforward:
\begin{equation}
\Phi=\Phi_c \frac{\sin \lambda r}{\lambda r}, \end{equation} where
\begin{equation}
\lambda^2=\frac{4\pi G}{\kappa^2}
\end{equation}
and $\Phi_c$ is constant. There is a particularly neat relation
for $\rho$ using equation (\ref{eq:neat}) yielding
\begin{equation}
4\pi G \rho + \frac{4\pi G}{\lambda^2} \Phi =0,
\end{equation}
which gives \begin{equation} \rho(r) = \rho_c \frac{\sin \lambda
r}{\lambda r}. \end{equation} The resulting density profile will
be recognized as that of an $n=1$ polytrope, i.e. a fluid in which
the gas pressure is related to the density via $p=K\rho^2$, with
$K$ a constant.

The important point emerging from this analysis is that one can
only obtain a static solution in the presence of the quantum
pressure term. A self-consistent solution requires this term to be
included and its effect is to stabilize the system. In order to
apply the wave mechanical approach to structure formation,
therefore, one has to be more careful with this term than Widrow
\& Kaiser (1993) in their equation (\ref{eq:schrod}). We
investigate this issue further in the following Section.

\section{One-dimensional Collapse}
In this section we look at some simple one-dimensional
 examples of gravitational instability using the Schr\"{o}dinger equation.
We do this by comparing the solutions we obtain with the
Zel'dovich approximation, which gives an exact solution up to the
point of shell-crossing (Shandarin \& Zel'dovich 1989). We also
want to consider the effects of varying the parameter $\nu$ in the
Schr\"{o}dinger equation and how this changes the spatial and
velocity resolution of the solution, as well as elucidating more
clearly the effects the ``quantum pressure'' term.

\subsection{The ``Free-Particle'' Solution to the Schr\"{o}dinger Equation}
\begin{figure*}
\begin{center}
\includegraphics[width=450pt,height=350pt]{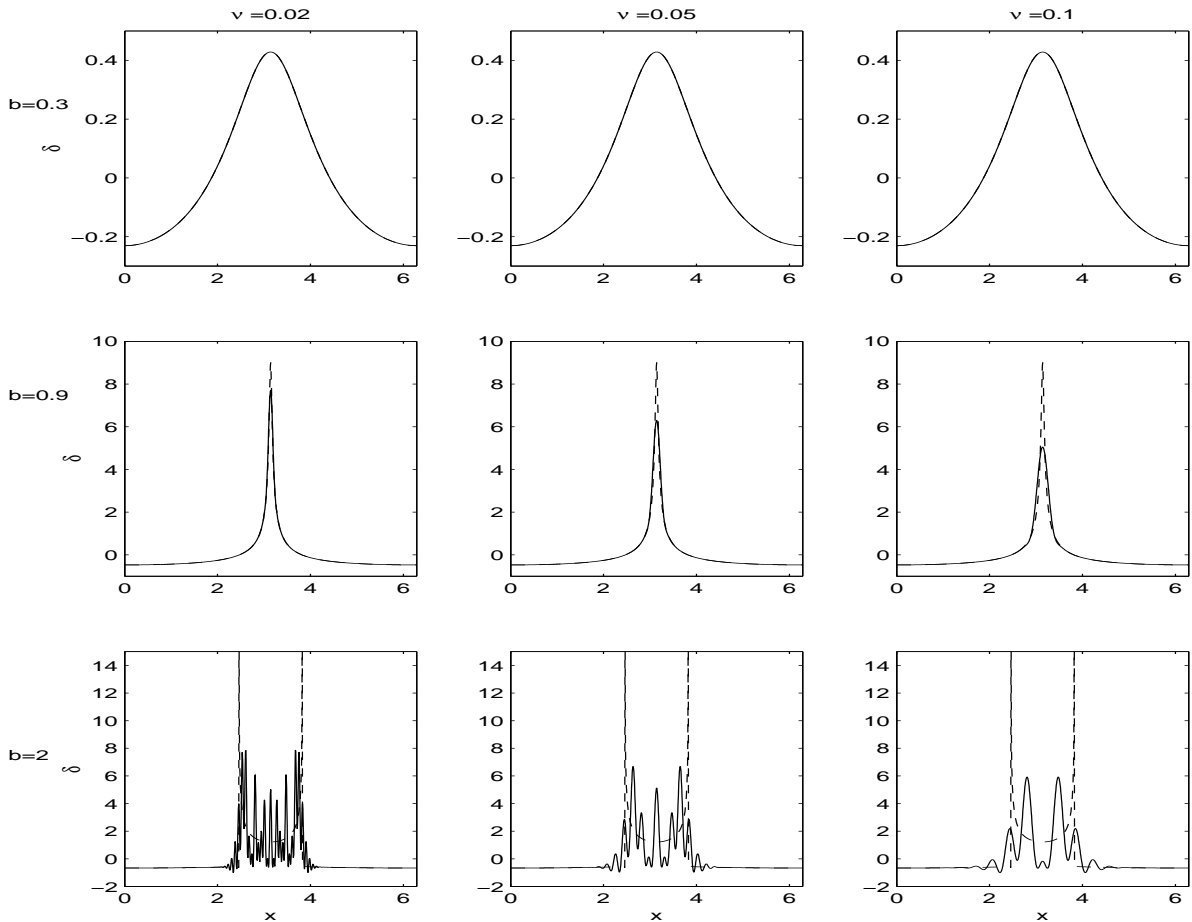}
\caption{Comparison of the Zel'dovich approximation to the
``free-particle'' Schr\"{o}dinger equation. The Zel'dovich
approximation is given by the dashed line and the solution to the
Schr\"{o}dinger equation by the solid line.  At early times the
two methods are almost identical. Close to shell crossing the
results are similar, but with the Schr\"odinger result behaving
more smoothly than the Zel'dovich solution. The effect of varying
$\nu$ can be most clearly seen after shell-crossing.}
\end{center}
\end{figure*}

The equation for the trajectory of a particle using the Zel'dovich
approximation is identical in form to that of inertial (or
ballistic) motion, with the displace replaced by the comoving
co-ordinate $x$ and the time replaced by $b$ as defined in
equation (22). The most obviously analogous approach to take with
the Schr\"{o}dinger equation is  obtained be setting the potential
$V$ equal to zero, replacing $t$ with $b$ and taking spatial
derivatives with respect to the comoving co-ordinate $x$. Equation
(31) then becomes
\begin{equation}
i\nu \frac{\partial \psi}{\partial b}=-\frac{\nu^2}{2}\nabla_{\bf x}^2\psi,
\end{equation}
or, in one-dimension,
\begin{equation}
i\nu \frac{\partial \psi}{\partial b}=-\frac{\nu^2}{2}\frac{\partial^2\psi}{\partial x^2}.
\end{equation}
This is the ``free-particle'' Schr\"{o}dinger equation and the
solutions are given by
\begin{equation}
\psi(x,t)=\sum_k A_k e^{(kx-\omega b)},
\end{equation}
with $ \omega=k^2\nu /2$. The boundary conditions are given by the
initial value of the wave function at $b=0$:
\begin{equation}
\psi(x,0)=\sum_k A_k e^{ikx}.
\end{equation}
The values of $A_k$ can be calculated using the Fast Fourier
Transform of the initial wavefunction, and the value of the
wavefunction at any future time can be calculated directly using
equation (47).

In order to show the equivalence between the Zel'dovich
approximation and Schr\"{o}dinger method we consider the two
solutions when the initial velocity potential is given by a single
sine wave. In  the Schr\"{o}dinger equation we start with an
initial wave function given by
\begin{equation}
\psi(x,0)=\sqrt{\rho_0}\exp(-i\cos x/\nu).
\end{equation}
This is equivalent to setting the initial density contrast
$\delta=0$ and the velocity $v=\sin x$. In one dimension, the
Zel'dovich approximation for the position of a particle at time
$t$ is given by,
\begin{equation}
x(q,t)=q-b\frac{d\Phi_0(q)}{dq},
\end{equation}
cf. equation (21). Conservation of mass is enforced by
\begin{equation}
\rho_0dq=\rho dx
\end{equation}
and therefore the density contrast is given by,
\begin{equation}
\delta=\left(\frac{\partial x}{\partial q}\right)^{-1}-1=\left[1-b\frac{d^2\Phi_0(q)}{dq^2}\right]^{-1}-1.
\end{equation}
If the velocity potential $\phi_0(q)=\cos q$ then the density contrast at time $t$ is then given by
\begin{equation}
\delta=(1+b\cos x)^{-1}-1
\end{equation}
and shell-crossing occurs when $b=1$ when the density contrast
becomes infinite. Figure 1 shows the density contrast obtained
using both of these methods at three different times and for three
different values of $\nu$. When $b=0.3$ density growth is still
approximately linear and the density contrast given by the
Schr\"{o}dinger and Zel'dovich methods are almost identical. When
$b=0.9$, just before shell-crossing, the density growth has become
highly non-linear. The very high density peaks seen in the
Zel'dovich approximation are not seen in the Shro\"{o}dinger
solution.

The value of $\nu$ corresponds to a ``de Broglie wavelength''
which gives the spatial resolution of the solution. The effects of
this can be seen most clearly when $b=2.0$ after shell-crossing
has occurred. Although the Schr\"{o}dinger and Zel'dovich methods
show qualitatively different behaviour after shell-crossing, this
is not physical, and  the overall behaviour of the results is in
any case similar, particularly for small values of $\nu$.

\subsection{The Schr\"{o}dinger equation with time-independent potential}

\begin{figure*}
\begin{center}
\includegraphics[width=450pt,height=350pt]{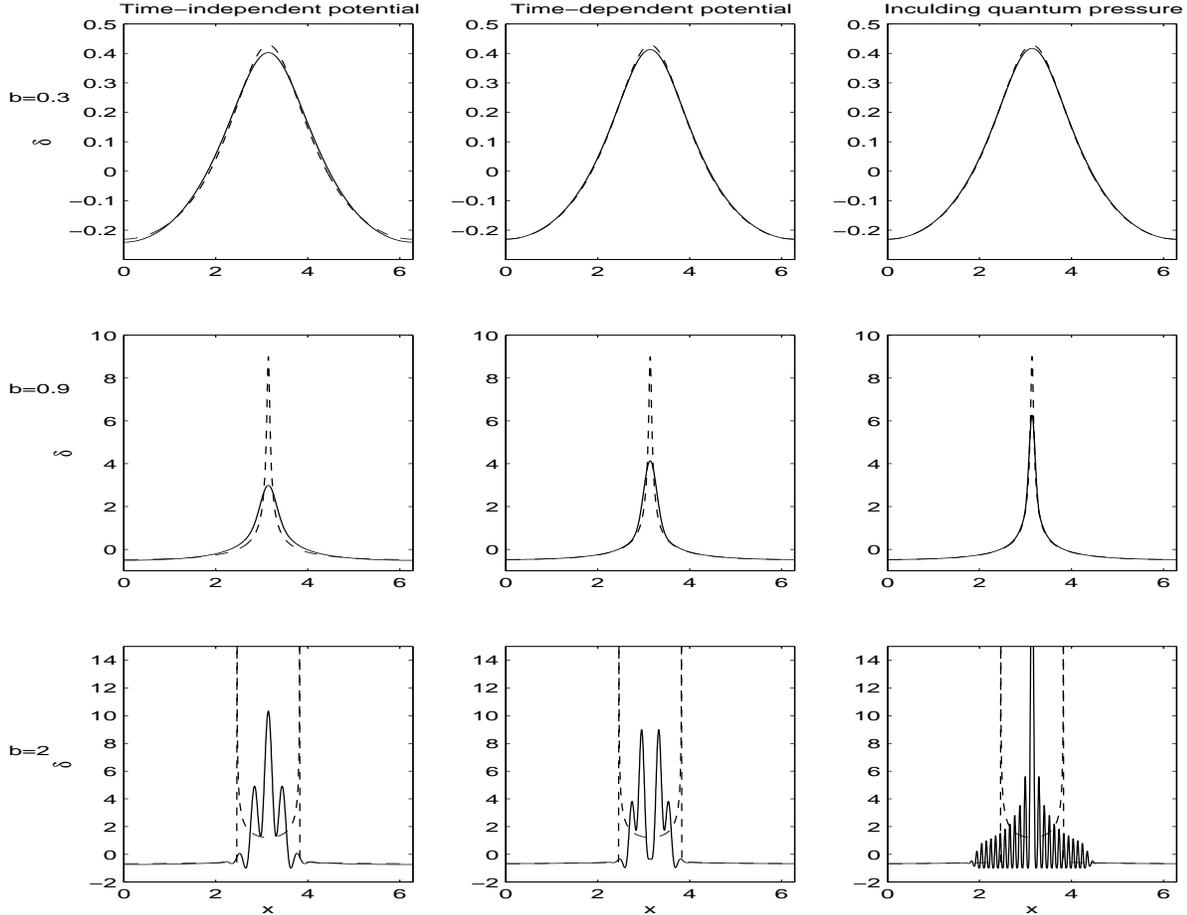}
\caption{Comparison of the Zel'dovich approximation to the
Schr\"{o}dinger equation with $\nu=0.1$.  In the first column the
potential $V$ does not vary with time.  In the second column
potential is calculated according to the Poisson equation and in
the third column the quantum pressure term is included.}
\end{center}
\end{figure*}

Setting the potential $V(x)=0$ in the Schr\"{o}dinger equation
gives a good agreement with the Zel'dovich approximation. However,
in both methods the particles continue on their trajectories after
shell-crossing with no reference to the self-gravity of the
collapsing system.  Here we look at how the inclusion of a
potential term in the Schr\"{o}dinger equation might give a more
realistic solution after shell-crossing. This approach is
motivated by the realisation that, at least during linear
perturbation growth, the gravitational potential corresponding to
a density fluctuation barely changes. In a different context this
has motivated the so-called ``frozen-potential'' approximation
(Matarrese et al. 1992; Brainerd, Scherrer \& Villumsen 1993;
Bagla \& Padmanabhan 1994), which is similar in spirit to the
approach we adopt here.

The simplest way to include the potential is to include its
initial value, but not to allow this to vary with time. For this
example we again ignore the quantum pressure term. We therefore
need to solve the equation
\begin{equation}
i\nu \frac{\partial\psi}{\partial t}=-\frac{\nu^2}{2}\frac{\partial^2 \psi}{\partial x^2}+\cos(x)\psi.
\end{equation}
It should be noted that in this example no account is taken of the
expansion of the universe. This means that a direct comparison
with this Zel'dovich approximation is not as straightforward as in
the previous example and in particular the time coordinate is not
the same in the two methods. However a comparison may be made by
considering the growth at early times in the linear regime.
 In the Zel'dovich approximation the density contrast is given by
\begin{equation}
\delta \simeq b\frac{\partial^2 \phi}{\partial x^2}\simeq -b\cos x,
\end{equation}
and the velocity in comoving coordinates is given by
\begin{equation}
v = \frac{{\rm d}x}{{\rm d}t}\simeq -\dot b \frac{{\rm d}\phi_0}{{\rm d} x}\simeq \dot b \sin x.
\end{equation}
In a non-expanding universe with a constant potential given by
$V(x)=V_0 \cos x$ the
 density contrast at early times is
\begin{equation}
\delta \simeq -\frac{V_0}{2} t^2 \cos x.
\end{equation}
We can therefore compare the Zel'dovich approximation at time $b$
with the Schr\"{o}dinger equation at time $t=\sqrt{2b/V_0}$.
Starting with the initial wavefunction given by \begin{equation}
\psi(x,0)=\sqrt{\rho_0},
\end{equation}
and evolving the wavefunction according to equation (54) using
Cayley's method (Goldberg, Schey \& Schwartz 1967) leads to the
results shown in the first column of Figure 2.  These are for
$\nu=0.1$ at times corresponding to the values of $b=0.3$, $0.9$
and $2.0$ used before. Comparison with Figure 1 shows that the
extensive smearing of density peaks which occurs after
shell-crossing in both the ``free-particle'' and Zel'dovich
solution is no longer seen in the case where the initial potential
is included.

\subsection{The Schr\"{o}dinger equation with time-varying potential}

In the second column of Figure 2 we show the solution to the
Schr\"{o}dinger equation with the full potential calculated using
the Poisson equation (33). We start be considering the equation
without the quantum pressure term. In this case we need to solve
the coupled Schr\"{o}dinger and Poisson equations:
\begin{equation}
i\nu \frac{\partial \psi}{\partial t}=-\frac{\nu^2}{2}\frac{\partial^2 \psi}{\partial x^2}+V(x)\psi;\end{equation}
\begin{equation}
\frac{{\partial}^2 V}{{\partial} x^2}=4 \pi G \psi \psi^*.
\end{equation}
To set up the initial wavefunction we choose a single sine wave for the density profile,
\begin{equation}
\delta=-\delta_0 \cos x.
\end{equation}
In the linear regime the density contrast at time $t$ is given by
\begin{equation}
\delta \simeq -\delta_0 \exp(\sqrt{4\pi G}t)\cos x
\end{equation}
and the velocity is
\begin{equation}
v=\sqrt{4\pi G}\delta_0\exp(\sqrt{4\pi G}t)\sin x.
\end{equation}
Accordingly we compare the Zel'dovich approximation at time $b$
with the Schr\"{o}dinger equation at time $t=1/\sqrt{4\pi G}\ln
(b/\delta_0)$. We start with the initial wavefunction,using
equation (35), given by,
\begin{equation}
\psi(x,0)=\sqrt{\rho_0(1-\delta_0 \cos x)}\exp \left(-i\sqrt{4\pi
G}\delta_0\cos x /\nu\right),
\end{equation}
and evolve the wavefunction according to equation (59). In this
case the potential $V(x)$ is calculated by solving the Poisson
equation numerically at each time step. The results are shown in
the second column of Figure 2 for $\nu=0.1$ at times corresponding
to $b=0.3$, $0.9$ and $2.0$ used before.

The third column of Figure 2 shows the results for the same
initial wavefunction, but this time including the quantum pressure
term, as in equation (31).  The inclusion of this term makes a
significant difference to the results at later times, particularly
after shell-crossing, but is slight earlier on during the
evolution. This means that useful approximations for the mildly
non-linear regime can still be be obtained even if this term is
ignored.

\subsection{Solutions in an expanding background}

Although we do not make further use of it in this paper, it is
useful to end this section with a few comments about what arises
when one places the system in an expanding background. To see what
happens, let us define a scaled density
$\chi=\rho/\rho_0=(1+\delta)$ and take $\Omega=1$. The continuity
equation then becomes
\begin{equation}
\frac{\partial \chi}{\partial a} + \nabla \cdot (\chi\nabla\phi)=
0~,
\end{equation}
where the velocity potential $\phi$ is now such that ${\bf
u}=d{\bf x}/dt=\nabla\phi$ and $a$ is the scale factor. It is
convenient to take the first integral of the Euler equation,
giving
\begin{equation}
\frac{\partial \phi}{\partial a} + \frac{1}{2} (\nabla \phi)^2 =
-\frac{3}{2a}(\phi+\theta),
\end{equation}
where $\theta=2\Phi/3a^3H^2$ and $\Phi$ is the gravitational
potential. After some more algebra the system again becomes a
Schr\"{o}dinger-like wave equation, but in $a$ rather than in $t$
and using $\psi^2=\chi$, such that
\begin{equation}
i\nu \frac{\partial \psi}{\partial a} = -\frac{\nu^2}{2} \nabla^2
\psi + V\psi + P\psi~, \label{eq:scrap2}
\end{equation}
with $V=\phi+\theta$ and $P$ as before. It is relatively
straightforward to incorporate the extra terms.

\section{Discussion}
In this paper we have set out an approach to the formation of
cosmic structure from evolving cosmological density fluctuations
that relies on a transformation of the evolution equations into a
Schr\"{o}dinger-Poisson system. One advantage of this new
formalism was pointed out by Coles (2002): it yields a rather more
convincing approach to understanding the origin of spatial
intermittency in the galaxy distribution than that offered by
Jones (1999). It also makes a connection in the underlying physics
with other systems that display similar phenomena. In this paper
we have argued further that it provides an analytical approach
that offers benefits over standard approaches of both Lagrangian
and Eulerian types. On the one hand, it always guarantees a
positive density and on the other it does not produce
singularities at shell-crossing events. It is clearly important to
investigate both of these issues further. Szapudi \& Kaiser (2003)
have already completed a study of perturbation theory and its
statistical ramifications within this approximation. Dealing fully
with post-shell-crossing phenomena is beyond the scope of this
paper as it requires a more sophisticated representation of the
wave-function in velocity space; see Widrow \& Kaiser (1993) for
discussion.

In this paper we showed how this approach works in practice by
applying it to a static spherically-symmetric case and to some
examples of one-dimensional collapse. In doing this we looked at a
``free-particle'' solution of the system that resembles the
Zel'dovich approximation in some respects. Incorporating a fixed
gravitational potential yields a solution that has some
similarities to the ``frozen-potential'' approximation (Matarrese
et al. 1992). We think the Schr\"{o}dinger formalism provides a
useful tool for modelling fluctuations in the mildly non-linear
regime. In future work we will look at its use in modelling the
intergalactic medium in order to understand properties of
Lyman-$\alpha$ absorption systems.

\section*{Acknowledgements}
We are grateful to Roya Mohayaee and Peter Watts for related
discussions.

\end{document}